\documentclass[preprint,prd,eqsecnum]{revtex4}
\usepackage{graphicx}
\usepackage{hyperref}
\usepackage{amsmath}
\usepackage{rcs}

\newcommand{\PRpt}{{\it Phys. Rep.}}

\newcommand{\PR}{{\it Phys. Rev.} }
\newcommand{\CQG}{{\it Class. Quant. Grav.}}
\newcommand{\MNRAS}{{\it Mon. Not. R. Astr. Soc.}}
\newcommand{\abs}[1]{\left|#1\right|}
\newcommand{\un}[1]{\mathrm{\,#1}}
\newcommand{\tens}[1]{\Vec{\Vec{#1}}}
\begin{document}
\title{Stochastic Gravitational Wave Measurements with Bar Detectors:
  \\ Dependence of Response on Detector Orientation}
\preprint{LIGO-P050041-01-Z}
\RCS $Date: 2005/09/27 05:07:16 $
\date{\RCSDate}
\author{John T Whelan}
\email{jtwhelan@loyno.edu}
\affiliation{Department of Physics, Loyola
    University, 6363 St Charles Ave Box 92, New Orleans, Louisiana
    70118, USA}

\begin{abstract}
  The response of a cross-correlation measurement to an isotropic
  stochastic gravitational-wave background depends on the observing
  geometry via the overlap reduction function.  If one of the
  detectors being correlated is a resonant bar whose orientation can
  be changed, the response to stochastic gravitational waves can be
  modulated.  I derive the general form of this modulation as a
  function of azimuth, both in the zero-frequency limit and at
  arbitrary frequencies.  Comparisons are made between pairs of nearby
  detectors, such as LIGO Livingston-ALLEGRO, Virgo-AURIGA,
  Virgo-NAUTILUS, and EXPLORER-AURIGA, with which stochastic
  cross-correlation measurements are currently being performed,
  planned, or considered.
\end{abstract}

\maketitle

\section{Background and motivation}

\subsection{Stochastic Backgrounds and Cross-Correlation Measurements}

One of the gravitational wave (GW) sources targeted by the current
generation of ground-based interferometric and resonant detectors is a
stochastic GW background (SGWB), produced by an
unresolved superposition of signals of astrophysical or cosmological
origin.\cite{Christensen:1992,Allen:1997,Maggiore:2000}
Direct measurements can be made of, or observational limits set on, a
SGWB by looking for correlations in the outputs of a pair of GW
detectors. 

The response of a pair of detectors to a SGWB depends in part on the
observing geometry, in the form of the relative orientation and
separation of the two detectors.
If one or both of the detectors is a resonant bar detector, its
orientation can conceivably be changed, thereby altering the observing
geometry and the response of the
cross-correlation experiment to a stochastic GW background.  For
example, the experimental setup of the ALLEGRO resonant bar detector
\cite{Mauceli:1996} allows for rotation of the detector.  This
produces a modulation of in the stochastic GW response which is
used in the cross-correlation measurements being conducted
between ALLEGRO and the interferometer (IFO) at the LIGO Livingston
Observatory (LLO).\cite{Finn:2001,Whelan:2003,Whelan:2005,Abbott:2006}

In the case of LLO and ALLEGRO, where the separation between the
detectors (about 40\,km) is small compared to the wavelength of
gravitational waves to which both detectors are sensitive (about
300\,km), the azimuth dependence of the response is very nearly the
same as for colocated detectors: sinusoidal with a period of
$180^\circ$.\cite{Finn:2001}.
This work considers analytically the general dependence of the
stochastic gravitational-wave response on the orientation of a bar
detector, and applies the results to existing pairs of detectors.

\subsection{Mathematical Details}

If a SGWB is
isotropic, unpolarized, Gaussian, and stationary, it is completely
described by its power spectrum.  It is conventional to express this
spectrum in terms of the GW contribution to the cosmological parameter
$\Omega=\rho/\rho_{\text{crit}}$:
\begin{equation}
  \label{eq:omegagw}
  \Omega_{\text{GW}}(f)=\frac{1}{\rho_{\text{crit}}}
  \frac{d\rho_{\text{GW}}}{d\ln f}=\frac{f}{\rho_{\text{crit}}}
  \frac{d\rho_{\text{GW}}}{df}
  \ .
\end{equation}
If $\tilde{h}_{1,2}(f)$ are the Fourier transforms of the
detector outputs, the correlation in the presence of a SGWB of
spectrum $\Omega(f)$ should be\cite{Allen:1999}
\begin{equation}
  \label{eq:ccexpect}
    \langle\widetilde{h}_1(f)^*
    \widetilde{h}_2(f')\rangle
    = \frac{3H_0^2}{20\pi^2}\delta(f-f')
    \gamma(|f|)\Omega_{\scriptstyle{\rm GW}}(|f|)
\end{equation}
where $\gamma(f)$ is a geometrical factor
\cite{Michelson:1987,Christensen:1992,Flanagan:1993}
involving the location and
orientation of the two detectors, which we describe in detail in
Section~\ref{s:overlap}.  This factor also
appears in the sensitivity of a standard cross-correlation
technique\cite{Allen:1999}, for example in the case of a background
whose $\Omega_{GW}(f)$ is constant over the frequency band of
interest:
\begin{equation}
  \label{eq:stochsens}
  \Omega_{\text{sens}}
  \sim
  \left(
    {T}
    \int df\frac{
      {\gamma^2(f)}}{f^6 {P_1(f) P_2(f)}}
  \right)^{-1/2}
  \ .
\end{equation}

\section{Definition of the overlap reduction function}
\label{s:overlap}

The factor $\gamma(f)$ appearing in \eqref{eq:ccexpect} and
\eqref{eq:stochsens} is typically (see e.g., \cite{Allen:1997})
defined as
\begin{equation}
  \label{eq:gammadef}
  \gamma(f) = \frac{5}{8\pi}\sum_{A=+,\times}\iint d^2\Omega_{\hat{n}}
  \,F_{1\,A}(\hat{n})\,F_{2\,A}(\hat{n})
  e^{i2\pi f\hat{n}\cdot(\vec{r}_1-\vec{r}_2)/c}
\end{equation}
where $\vec{r}_1$ and $\vec{r}_2$ are the positions of the two
detectors, and $F_{1\,A}(\hat{n})$ and $F_{2\,A}(\hat{n})$ are their
beam patterns.  We follow the lead of \cite{Flanagan:1993} by
factoring the response tensors (cf.\ Appendix~\ref{app:notat})
out of the integral and writing
\begin{equation}
  \gamma(f) = d_{1\,ab}\,\Gamma^{ab}_{cd}(\alpha,\hat{s})\,d_2{}^{cd}
\end{equation}
where
\begin{equation}
  \hat{s}=\frac{\vec{r}_1-\vec{r}_2}{\abs{\vec{r}_1-\vec{r}_2}}
\end{equation}
is the unit vector pointing from one detector to the other, and
\begin{equation}
  \alpha = \frac{2\pi f\abs{\vec{r}_1-\vec{r}_2}}{c}
\end{equation}
By using the definition of the transverse, traceless projector in
\eqref{eq:TTproj}, we note that we can write
\begin{equation}
  \label{eq:Gammaint}
  \Gamma^{ab}_{cd}(\alpha,\hat{s})
  = \frac{5}{4\pi} \iint d^2\Omega_{\hat{n}}\,
  P^{\text{TT}\hat{n}}{}^{ab}_{cd}\,e^{i\alpha\hat{n}\cdot\hat{s}}
\end{equation}
Note that this gives an alternate definition of the overlap reduction
function which is manifestly independent of any polarization basis:
\begin{equation}
  \label{eq:gammaalt}
  \gamma(f)=d_{1\,ab}\, d_2{}^{cd}\,
  \frac{5}{4\pi} \iint d^2\Omega_{\hat{n}}\,
  P^{\text{TT}\hat{n}}{}^{ab}_{cd}\,
  e^{i2\pi f\hat{n}\cdot(\vec{r}_2-\vec{r}_1)/c}
\end{equation}
Written in the form \eqref{eq:Gammaint}, it is clear that
$\Gamma^{ab}_{cd}(\alpha,\hat{s})$ is symmetric and traceless on both
pairs of indices ($\{ab\}$ and $\{cd\}$).  With $\hat{s}$ as the only
preferred direction, there are only three independent tensors which
can be created with these properties:
\begin{subequations}
  \begin{align}
    T_1{}^{ab}_{cd} &= P^{\text{T}}{}^{ab}_{cd} \\
    T_2{}^{ab}_{cd}(\hat{s})
    &= P^{\text{T}}{}^{ab}_{ef} \hat{s}^f \hat{s}_g P^{\text{T}}{}^{eg}_{cd} \\
    T_3{}^{ab}_{cd}(\hat{s})
    &= P^{\text{T}}{}^{ab}_{ef} \hat{s}^e \hat{s}^f \hat{s}_g \hat{s}_h P^{\text{T}}{}^{gh}_{cd}
  \end{align}
\end{subequations}
(Previous derivations \cite{Flanagan:1993,Allen:1997} included two
additional terms which were not traceless, the co\"{e}fficients of
which are of course zero.)  The most general possible form is thus
\begin{equation}
  \label{eq:Gammasum}
  \Gamma^{ab}_{cd}(\alpha,\hat{s})
  = \sum_{n=1}^3 \rho_n(\alpha) T_n{}^{ab}_{cd}
\end{equation}
and hence
\begin{equation}
  \label{eq:gammaresult}
  \gamma(f) = \rho_1(\alpha) d_1^{\text{T}}{}^{ab} d_2^{\text{T}}{}_{ab}
  + \rho_2(\alpha) d_1^{\text{T}}{}^{ab} \hat{s}_b \hat{s}^c d_2^{\text{T}}{}_{ac}
  + \rho_3(\alpha) d_1^{\text{T}}{}^{ab} \hat{s}_a \hat{s}_b
  \hat{s}^c \hat{s}^d d_2^{\text{T}}{}_{cd}
\end{equation}

The complete functional forms of the co\"{e}fficients are derived in
\cite{Flanagan:1993} and corrected in
\cite{Allen:1997,Christensen:1997}, and a slightly simplified
derivation appears in Appendix~\ref{app:deriv}
of the present paper.
First, however, it is
elucidating to note the behavior at $\alpha=0$, which corresponds
either to the low-frequency limit or to the case where the two
detectors are co-located.  In that case, \eqref{eq:Gammaint} becomes
\begin{equation}
  \Gamma^{ab}_{cd}(0,\hat{s})
  = \frac{5}{4\pi} \iint d^2\Omega_{\hat{n}}\,
  P^{\text{TT}\hat{n}}{}^{ab}_{cd}
\end{equation}
which is manifestly independent of the separation direction $\hat{s}$.
That tells us that it can contain only the $\hat{s}$-independent term
$T_1{}^{ab}_{cd}$.  I.e., $\rho_2(0)=0=\rho_3(0)$, leaving
\begin{equation}
  \label{eq:Gamma0}
  \Gamma^{ab}_{cd}(0,\hat{s})
  = \rho_1(0) P^{\text{T}}{}^{ab}_{cd}
  = \frac{5}{4\pi} \iint d^2\Omega_{\hat{n}}\,
  P^{\text{TT}\hat{n}}{}^{ab}_{cd}
\end{equation}
It's easy to solve for $\rho_1(0)$ by taking the trace of
\eqref{eq:Gamma0} and using \eqref{eq:PTTtrace} and \eqref{eq:PTtrace}
to say
\begin{equation}
  \Gamma^{ab}_{ab}(0,\hat{s})
  = \rho_1(0)\cdot 5
  = \frac{5}{4\pi} \iint d^2\Omega_{\hat{n}}\cdot 2 = 10
\end{equation}
This means that $\rho(0)=2$ and
\begin{equation}
  \label{eq:gamma0}
  \gamma(0) = 2 d_1^{\text{T}}{}^{ab} d_2^{\text{T}}{}_{ab}
\end{equation}
Note that this makes, for instance, the demonstration that the overlap
reduction function for co\"{\i}ncident, co\"{a}ligned IFOs
with perpendicular arms is unity, extremely simple.  [Just substitute
\eqref{eq:difo} in for both response tensors.]

\section{Dependence of the overlap reduction function on
  the orientation of a bar detector}

In this section we consider in detail the form of the overlap
reduction function between a bar detector with its long axis along the
unit vector $\hat{u}$ and
\begin{enumerate}
\item an IFO with axes along the unit vectors $\hat{x}$ and
  $\hat{y}$
\item a bar detector with its long axis along the unit vector $\hat{x}$
\item a general detector with response tensor $d_{ab}$
\end{enumerate}

\subsection{Zero-frequency limit}

The relatively simple form \eqref{eq:gamma0} of $\gamma(0)$ makes it
worthwhile to consider briefly the overlap reduction function in this
limit, which can in general be written as
\begin{equation}
  \frac{2\pi f\abs{\vec{r}_1-\vec{r}_2}}{c} \rightarrow 0
\end{equation}
and thus also applies to detectors located at the same site.

\subsubsection{Correlations with an interferometer}

For correlations between an IFO and a bar, we use
\eqref{eq:difo} and \eqref{eq:dbar} and note that in this case
$d_1^{ab}$ is already traceless, so
\begin{equation}
  \gamma(0) = 2 d_1^{ab}\, d_{2\,ab}
  = \frac{1}{2}[(\hat{x}\cdot\hat{u})^2 - (\hat{y}\cdot\hat{u})^2]
\end{equation}
For the case where the detectors lie in the same plane, the
IFO's arms are perpendicular, and the bar makes an angle of
$\theta$ with the IFO's ``x arm'' and
$\frac{\pi}{2}-\theta$ with its ``y arm'', so that
\begin{subequations}
  \begin{align}
    \hat{x}\cdot\hat{u} &= \cos\theta \\
    \hat{y}\cdot\hat{u} &= \sin\theta    
  \end{align}
\end{subequations}
we have the familiar result that
\begin{equation}
  \label{eq:gamma0ifobar}
  \gamma(0) = \cos 2\theta
\end{equation}

\subsubsection{Correlations with another bar}
\label{sss:modulation-DC-barbar}

In the case of correlations between two bar detectors, the response
tensors are
\begin{subequations}
  \begin{align}
      d_1^{ab} &= \hat{x}^a \hat{x}^b \\
      d_2^{ab} &= \hat{u}^a \hat{u}^b
  \end{align}
\end{subequations}
so that from \eqref{eq:gamma0}
\begin{equation}
  \label{eq:gamma0bars}
  \gamma(0) = 2 
  \left(
    d_1^{ab} d_{2\,ab}
    - \frac{1}{3} d_1{}^a_a d_2{}^b_b
  \right)
  = 
  2 \left( (\hat{x}\cdot\hat{u})^2 - \frac{1}{3} \right)
  = \cos 2\theta + \frac{1}{3}
\end{equation}
where $\theta$ is the angle between the two bars, so that once again
$\hat{x}\cdot\hat{u} = \cos\theta$.

Note that the maximum value of \eqref{eq:gamma0bars}, which occurs
when $\theta=0$, is $\frac{4}{3}$, as opposed to the maximum of unity
for correlations between an IFO and a bar \eqref{eq:gamma0ifobar} or
two IFOs.  For this reason, the overlap reduction function for two
bars is sometimes (see, e.g., \cite{Maggiore:2000,Michelson:1987})
normalized with an
additional factor of $\frac{3}{4}$ relative to \eqref{eq:gammadef} so
that the maximum at zero frequency is zero for any type of detector
pair.  The normalization used in this paper is preferred, however, for
several reasons:
\begin{itemize}
\item Including a detector-dependent normalization factor in the
  definition of $\gamma(f)$ would mean formulas like
  \eqref{eq:stochsens} would have different forms for different types
  of detector pairs.
\item While the maximum value of $\gamma(0)$ for two bars is
  $\frac{4}{3}$ rather than 1, its minimum value is $-\frac{2}{3}$
  rather than $-1$ (which is the minimum value for either two IFOs or
  an IFO and a bar).  So in all three cases the amplitude of the
  orientation-induced modulation is 1.
\end{itemize}
The latter point can be physically understood in terms of a bar
detector being ``more omnidirectional'' than an IFO.  For
an IFO, there is one optimal propagation direction for
which the GW response is a maximum, namely a wave propagating
perpendicular to the IFO plane.  Two IFOs with their arms
parallel will both respond ideally to waves propagating in this
direction.  The same is true if we rotate one of the IFOs $90^\circ$,
except that the complete correlation of the signals has become an
anticorrelation, so $\gamma = -1$.  On the other hand, a bar detector
has a one-parameter family of ``optimal'' propagation directions,
all perpendicular to the bar.  Two parallel bars share the same family
of optimal directions.  However, if we rotate one bar $90^\circ$, the
planes perpendicular to the bars no longer co\"{\i}ncide, and there is
only a single optimal propagation direction perpendicular to the bars
for which they both respond optimally.  So the anticorrelation in the
perpendicular arrangement is not as efficient as the correlation in
the parallel arrangement, and $\abs{\gamma(0)}$ is smaller for
perpendicular bars than for parallel ones.

\subsection{General form}

Returning to consideration of the overlap reduction function
$\gamma(f)$ for general frequencies and detector separations, what
interests us here is the dependence of $\gamma(f)$ on the orientation
of the bar.  We assume that the locations of the two detectors and the
geometry of the other detector are fixed.  The one variable is the
azimuth of the bar, which is conventionally defined as an angle
measured clockwise from the local geographic North.  If we define unit
vectors $\hat{N}$ and $\hat{E}$ pointing North and East, respectively,
the orientation vector for a bar with azimuth $\zeta$ is
\begin{equation}
  \hat{u} = \hat{N}\cos\zeta + \hat{E}\sin\zeta
\end{equation}
its response tensor is
\begin{equation}
  d_2^{ab} = (\hat{N}^a\cos\zeta + \hat{E}^a\sin\zeta)
  (\hat{N}^b\cos\zeta + \hat{E}^b\sin\zeta)
  = d_0^{ab} + d_C^{ab}\cos 2\zeta + d_S^{ab}\sin 2\zeta 
\end{equation}
where
\begin{subequations}
  \begin{align}
    d_0^{ab} &= \frac{\hat{N}^a \hat{N}^b+\hat{E}^a \hat{E}^b}{2} \\
    d_C^{ab} &= \frac{\hat{N}^a \hat{N}^b-\hat{E}^a \hat{E}^b}{2} \\
    d_S^{ab} &= \frac{\hat{N}^a \hat{E}^b+\hat{E}^a \hat{N}^b}{2}
  \end{align}
\end{subequations}
This decomposition of the response tensor into a piece independent of
the azimuth, plus pieces proportional to the sine and cosine of twice
the azimuth, allows us to write the overlap reduction function as
\begin{equation}
  \label{eq:gammadecomp}
  \gamma(f) = \gamma_0(f) + \gamma_C(f)\cos 2\zeta
  + \gamma_S(f)\sin 2\zeta
\end{equation}
where
\begin{equation}
  \gamma_{0,C,S}(f) = d_{1\,ab}\,\Gamma^{ab}_{cd}(\alpha,\hat{s})
  \,d_{0,C,S}^{cd}
\end{equation}
Given the location of a bar, it is relatively straightforward to work
out the components in some convenient Cartesian basis of the unit
vectors $\hat{N}$ and $\hat{E}$ at the location of the bar,
and thus construct the tensors
$d_0^{ab}$, $d_C^{ab}$, and $d_S^{ab}$.  Those can be used, along with
the information about the geometry of the other detector and the
separation between the two detectors, to construct the
co\"{e}fficients $\gamma_0(f)$, $\gamma_C(f)$, and $\gamma_S(f)$.

One final improvement on the form \eqref{eq:gammadecomp} is to define
$\gamma_A(f)$ and $\zeta_{\text{max}}(f)$ according to
\begin{subequations}
  \begin{align}
    \gamma_C(f) &= \gamma_A(f) \cos 2\zeta_{\text{max}}(f) \\
    \gamma_S(f) &= \gamma_A(f) \sin 2\zeta_{\text{max}}(f)  
  \end{align}
\end{subequations}
so that
\begin{equation}
  \label{eq:gammafinal}
  \gamma(f) = \gamma_0(f)
  + \gamma_A(f)\cos 2(\zeta - \zeta_{\text{max}}(f))
\end{equation}
At a given frequency $f$, $\gamma_0(f)$ is the orientation-independent
piece of the overlap reduction function, $\gamma_A(f)$ is the
amplitude of the orientation-dependent modulation, and
$\zeta_{\text{max}}(f)$ is the orientation (modulo $\pi$) for which
the overlap reduction function is a maximum.  Note that this is not
necessarily the \emph{optimal} alignment for stochastic background
observations; if $\gamma_0(f)<0$ at the frequency of interest, the
stochastic sensitivity is maximized by setting the azimuth to
$\zeta_{\text{max}}(f) + \frac{\pi}{2}$ (again modulo $\pi$).

\subsection{Examples for real-world detectors}

\label{ss:modulation-examples}

This section contains several applications of \eqref{eq:gammafinal} to
pairs of real-world detectors.  In each case the \texttt{detgeom}
suite of matlab routines \cite{detgeom} was used to calculate the
orientation-independent offset $\gamma_0(f)$, the amplitude
$\gamma_A(f)$ of the modulation and the azimuth
$\zeta_{\text{max}}(f)$ of maximum overlap, as a function of
frequency.

The major bar detectors around the world were oriented roughly
parallel to one another as part of agreements through 
the International Gravitational
Event Collaboration (IGEC).  This orientation is referred to here as
``the IGEC orientation''.

\begin{figure}[htbp]
  \begin{center}
    \includegraphics[width=3in,angle=0]{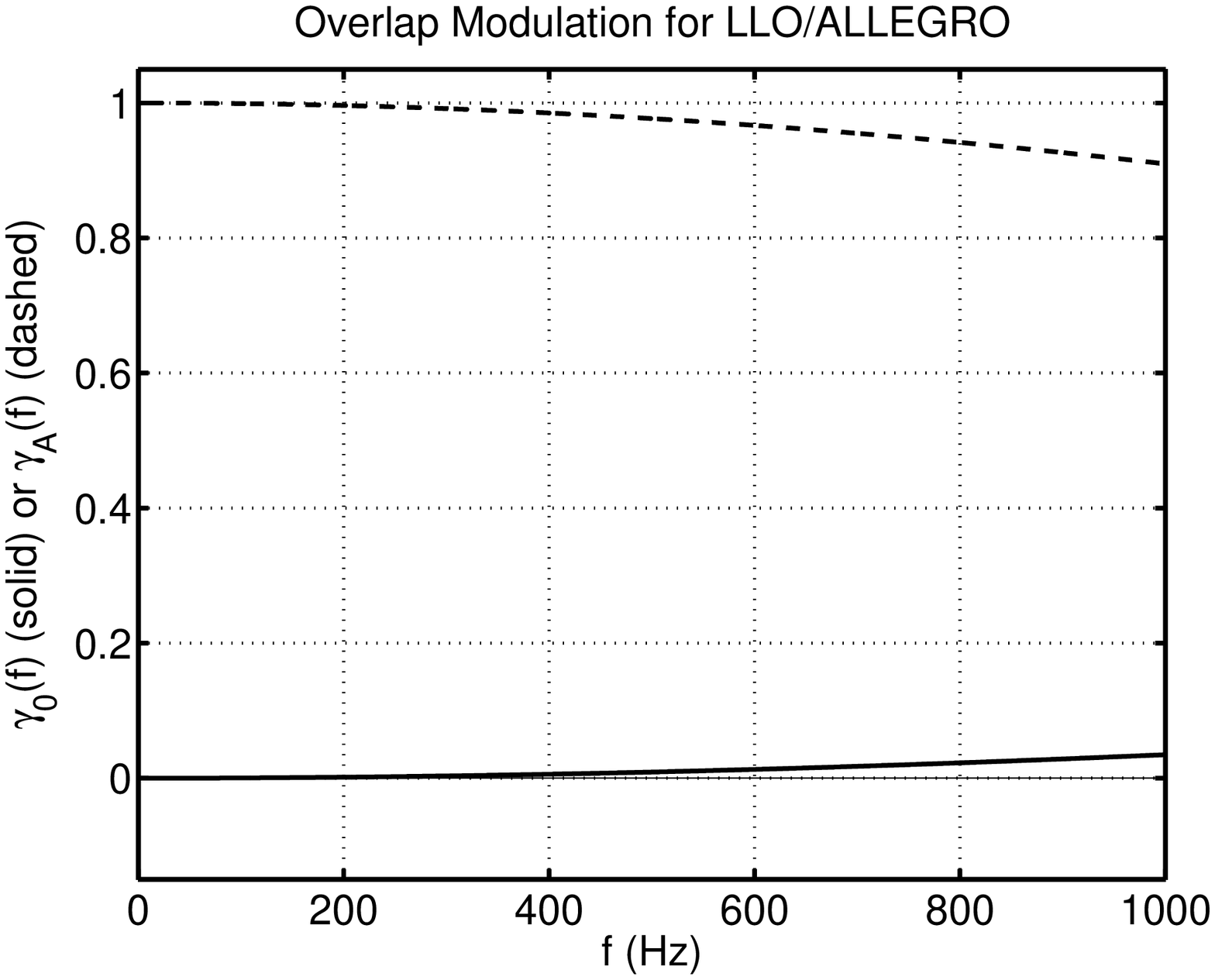}
    \includegraphics[width=3in,angle=0]{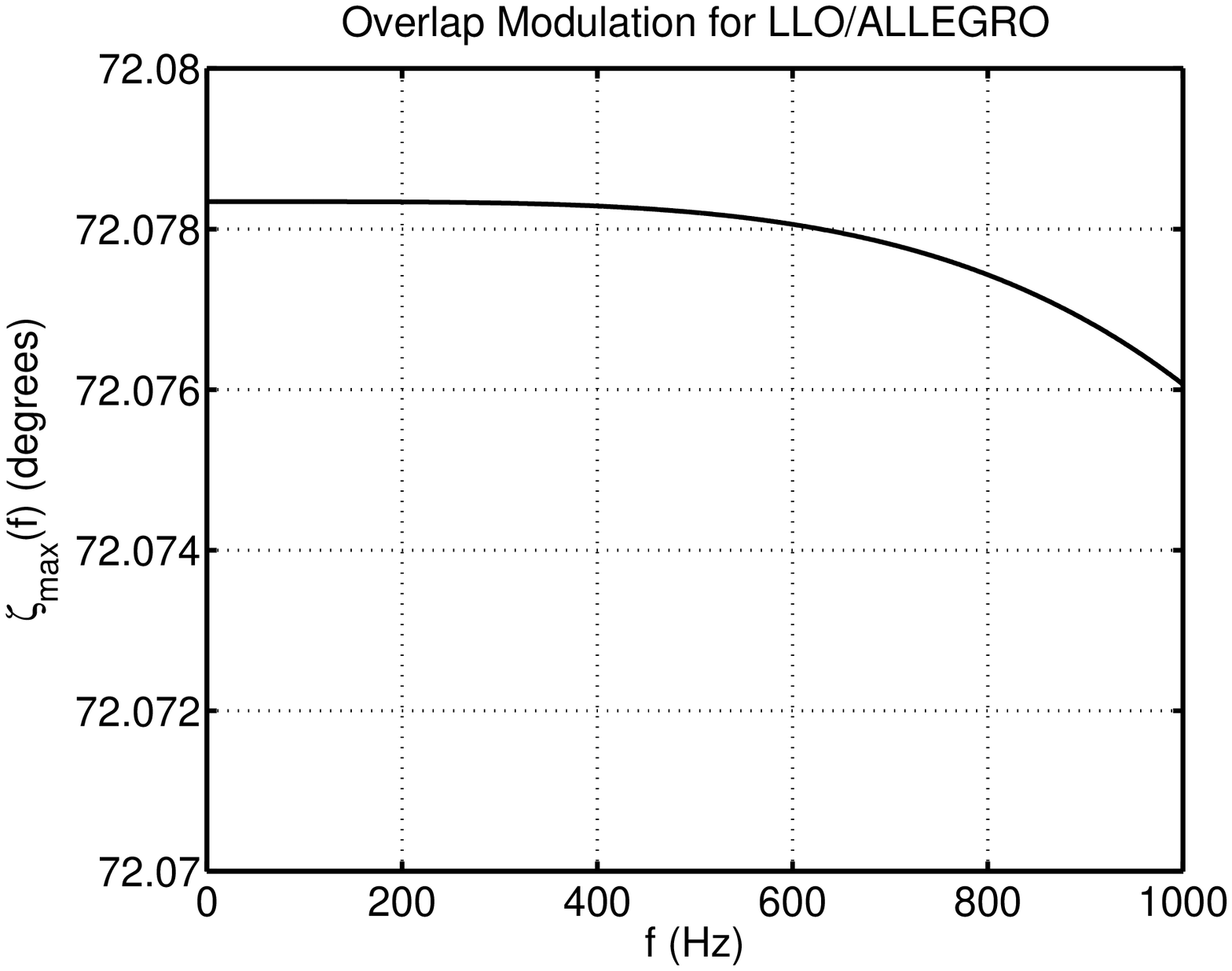}
  \end{center}  
  \caption{Modulation of the overlap reduction function for the
    ALLEGRO bar correlated with the LIGO Livingston interferometer.
    In the plot at left, the dashed line shows the amplitude
    $\gamma_A(f)$ of the azimuth-dependent oscillation, while the solid
    line indicates the azimuth-independent component $\gamma_0(f)$.
    The detector sites are only separated by 40\,km, so the amplitude
    and offset of the modulation change little between 0\,Hz and
    1000\,Hz.  The sensitive frequencies for the correlation measurement
    are near 900\,Hz, and the range of overlap reduction function
    values as the orientation is changed is
    $-0.90<\gamma(900\un{Hz})<0.96$, which makes this pair of detectors
    well suited to the modulation of stochastic GW response.
    Likewise, the optimal orientation differs from that
    determined at zero frequency by only a fraction of a degree.  The
    IGEC orientation, not shown on these axes, is -40, i.e., $40^\circ$
    West of North.}
  \label{fig:LLOALLEGRO}
\end{figure}
Fig.~\ref{fig:LLOALLEGRO} shows the modulation for the ALLEGRO bar
detector (Baton Rouge, LA, USA) correlated with the LIGO Livingston
interferometer (Livingston, LA, USA).  This is the closest pair of
detector sites, separated by only 40\,km.

\begin{figure}[htbp]
  \begin{center}
    \includegraphics[width=3in,angle=0]{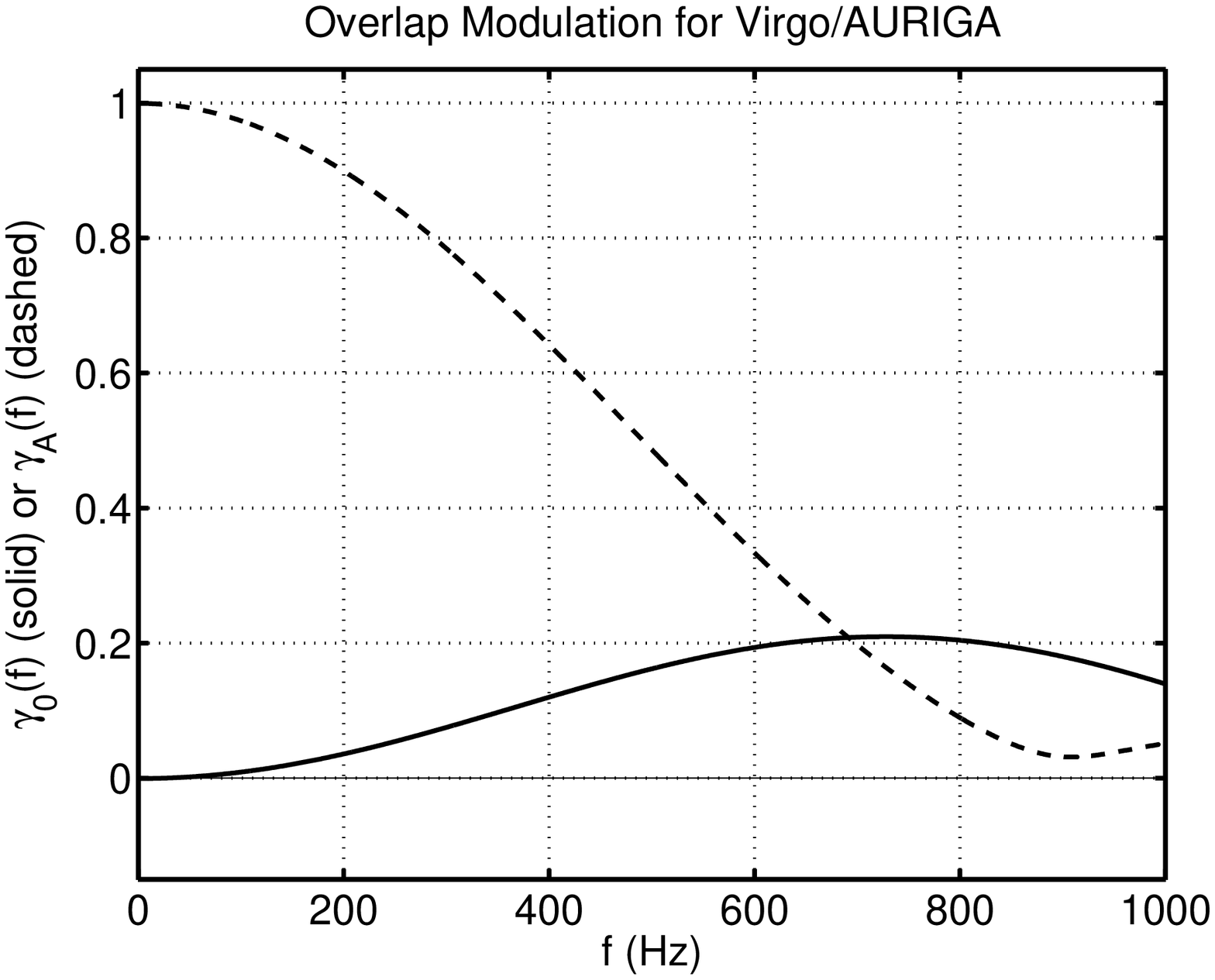}
    \includegraphics[width=3in,angle=0]{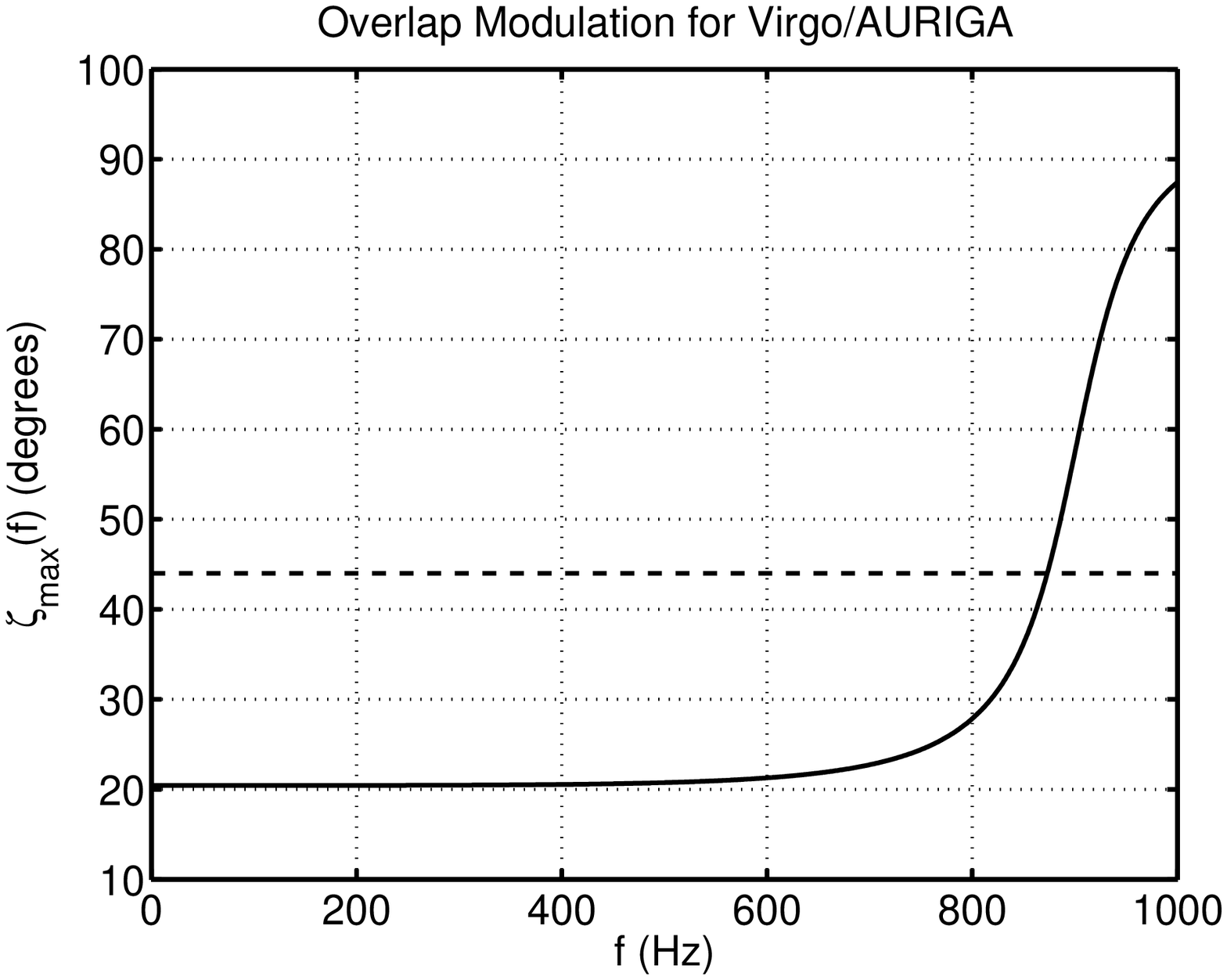}
  \end{center}  
  \caption{Modulation of the overlap reduction function for the AURIGA
    bar correlated with the Virgo interferometer.  The dashed line on
    the right-hand plot indicates AURIGA's IGEC orientation, for
    reference.  The detector sites are separated by 220\,km, so the
    frequency dependence of the overlap reduction function is
    important.  It happens that in the azimuthal modulation
    $\gamma_A(f)$ is rather small at the frequencies of the bar's
    sensitivity (around 900\,Hz).  However, the azimuth-independent
    offset is non-negligible, which means that
    $0.14<\gamma(900\un{Hz})<0.22$ for all orientations.  In
    particular, it is not possible to change the sign of the
    gravitational-wave response by altering the orientation of a
    detector at the AURIGA site.}
  \label{fig:VIRGOAURIGA}
\end{figure}
Fig.~\ref{fig:VIRGOAURIGA} examines the modulation for the AURIGA bar
(Legnaro, Italy) correlated with the Virgo interferometer (Cascina,
Italy).  AURIGA is the closest bar detector to Virgo, separated by
220\,km.

\begin{figure}[htbp]
  \begin{center}
    \includegraphics[width=3in,angle=0]{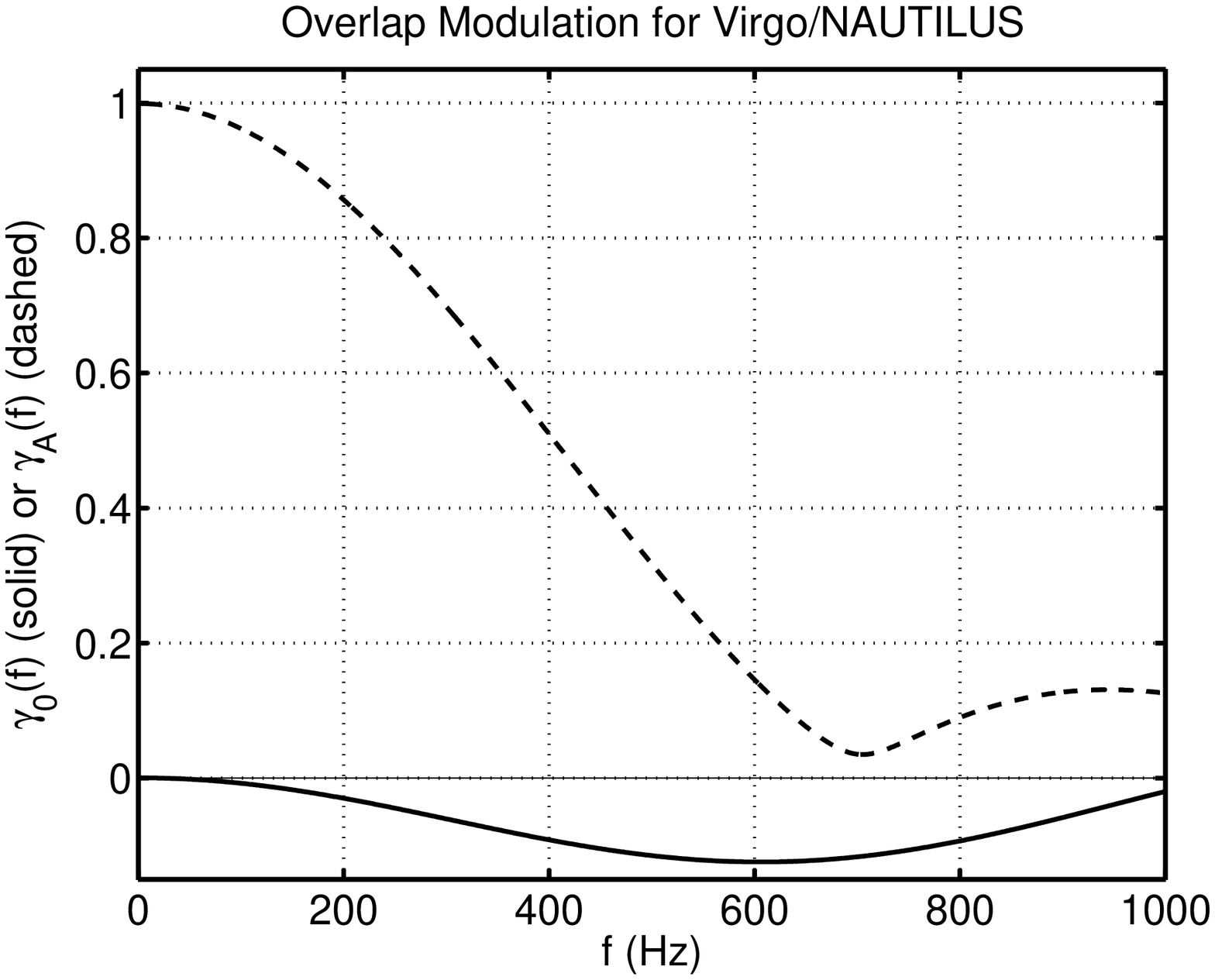}
    \includegraphics[width=3in,angle=0]{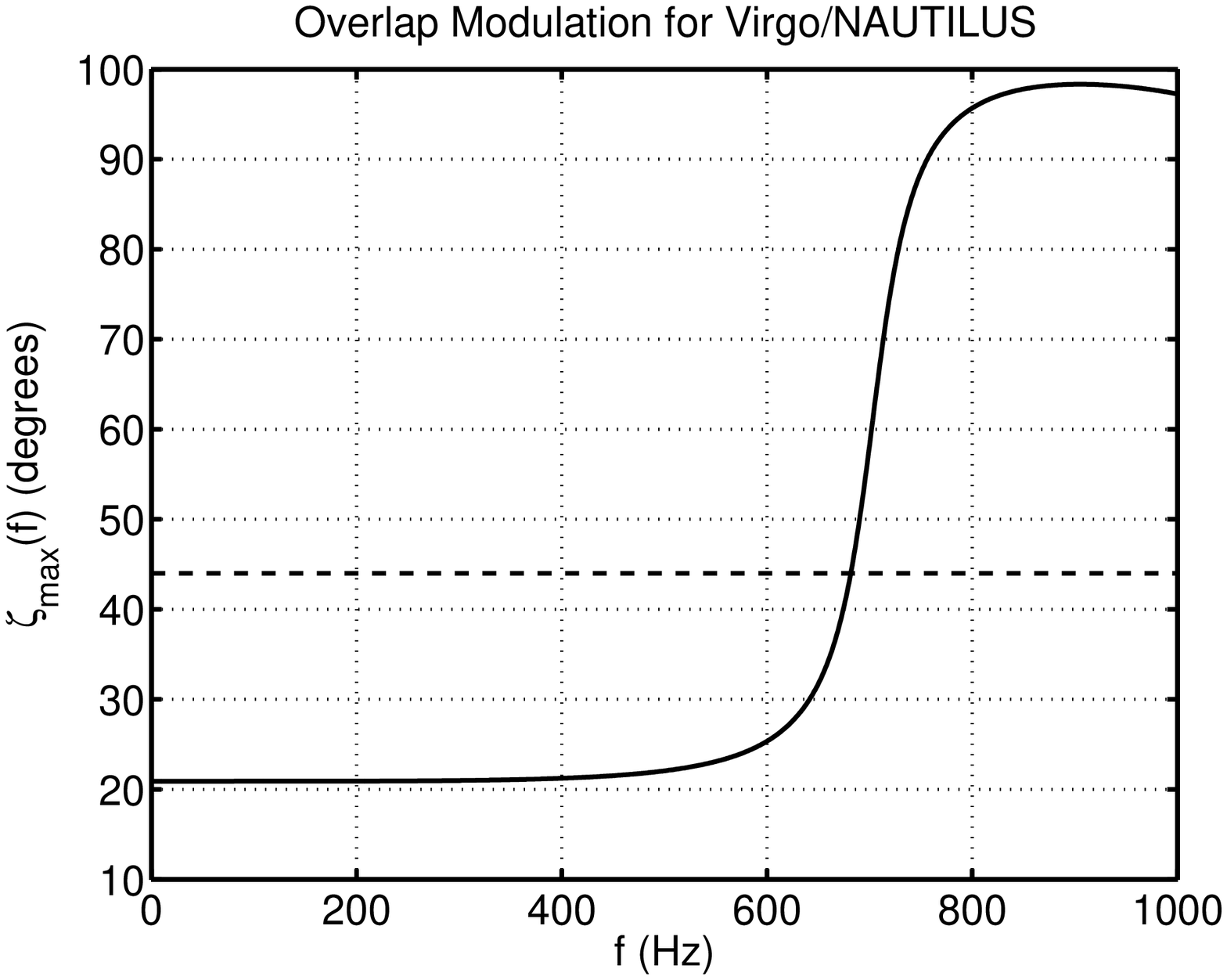}
  \end{center}  
  \caption{Modulation of the overlap reduction function for the
    NAUTILUS bar correlated with the Virgo interferometer.  The dashed
    line on the right-hand plot indicates NAUTILUS's IGEC orientation,
    for reference.  Since the detectors are separated by 270\,km, the
    amplitude $\gamma_A(f)$ of the modulation has its first minimum
    below 900\,Hz, and the range of overlap reduction function values
    is $-0.18<\gamma(900\un{Hz})<0.07$, depending on the bar's azimuth
    angle.  Note also that while
    $\zeta_{\text{max}}(900\un{Hz})=98^\circ$, this is the azimuth for
    which $\gamma(900\un{Hz})$ is a maximum; the maximum of
    $\abs{\gamma(900\un{Hz})}$ is in the perpendicular alignment, an
    azimuth of $8^\circ$ East of North.  Note also that while the
    actual IGEC azimuth of $44^\circ$ is $36^\circ$ away from this
    optimal orientation, this is not as close to a ``null'' alignment
    as it seems because of the offset $\gamma_A(900\un{Hz})=-0.06$,
    which makes $\gamma(900\un{Hz})=-0.10$ for the IGEC orientation.}
  \label{fig:VIRGONAUTILUS}
\end{figure}
Fig.~\ref{fig:VIRGONAUTILUS} shows the modulation for the NAUTILUS bar
(Frascati, Italy) correlated with the Virgo interferometer.  The two
sites are separated by 270\,km.

\begin{figure}[htbp]
  \begin{center}
    \includegraphics[width=3in,angle=0]{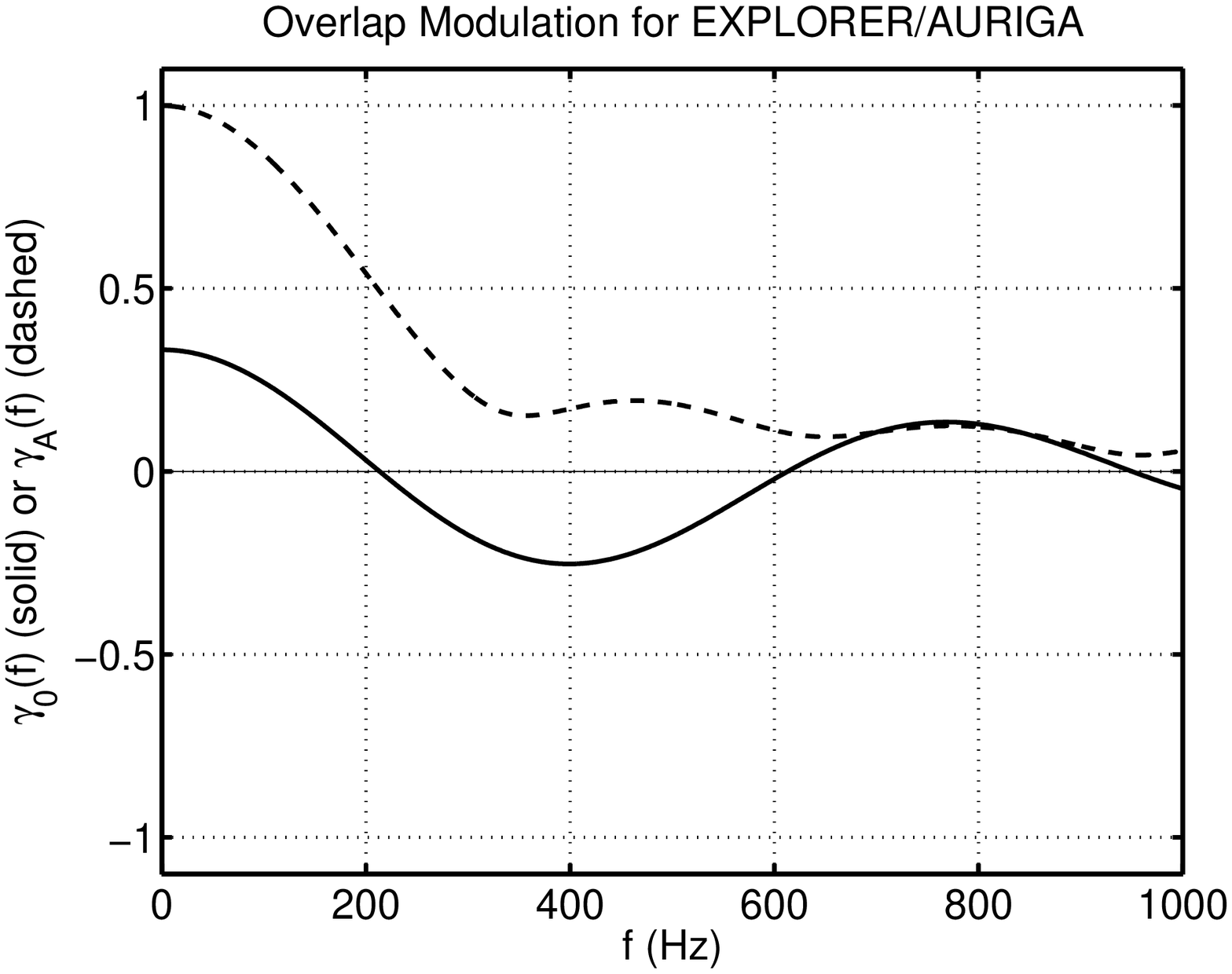}
    \includegraphics[width=3in,angle=0]{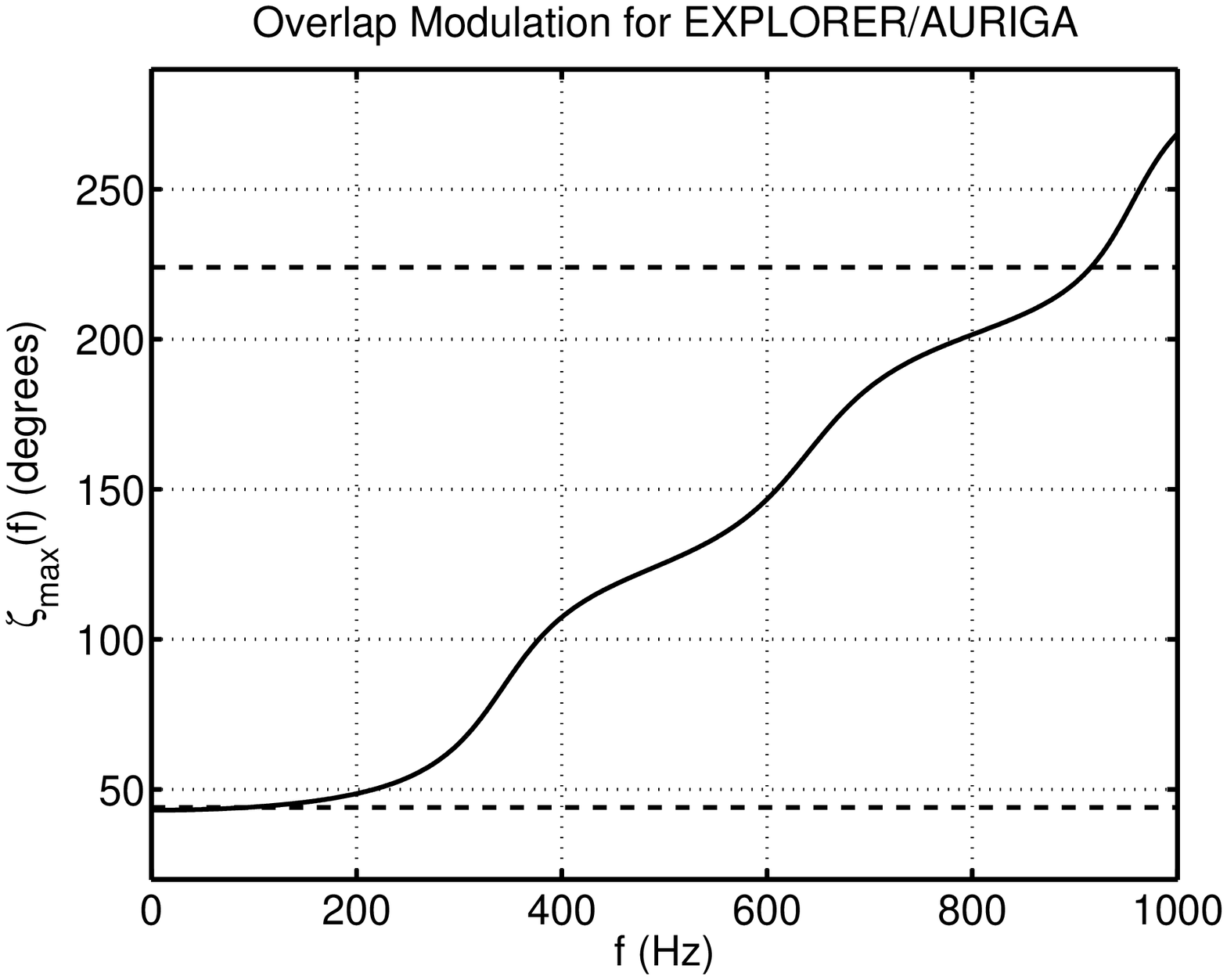}
  \end{center}  
  \caption{Modulation of the overlap reduction function for the AURIGA
    bar correlated with the EXPLORER bar.  EXPLORER is assumed to be
    in its actual IGEC orientation of 39$^\circ$ East of North.  Note
    that, as discussed in Sec.~\protect\ref{sss:modulation-DC-barbar},
    the offset of the modulation at zero frequency is $1/3$ rather
    than zero.  Note also that at zero frequency, the azimuth
    $\zeta_{\text{max}}(0)$ of maximum overlap is nearly equal to the
    actual IGEC azimuth, which is to be expected since in that
    orientation the bars are nearly parallel.}
  \label{fig:EXPLORERAURIGA}
\end{figure}
The result \eqref{eq:gammafinal} can also be applied to a pair of bar
detectors.  Fig.~\ref{fig:EXPLORERAURIGA} shows the modulation for the
AURIGA bar detector correlated with
the EXPLORER bar (Geneva, Switzerland).  The detectors are separated
by 460\,km.  EXPLORER is assumed to be in its actual IGEC orientation
of 39$^\circ$ East of North, and the modulation is that associated
with the AURIGA azimuth.

\section{Conclusions}

For stochastic background measurements correlating a resonant bar
detector with another GW detector, we have derived several analytical
results about the dependence of the overlap reduction function
$\gamma(f)$, and hence the response to an isotropic SGWB, on the bar's
orientation.  In particular:

At zero frequency, the overlap reduction function for any pair of
detectors is twice the contraction of the traceless parts of the
detectors' response tensors \eqref{eq:gamma0}.  For correlations 
between an interferometer and a bar lying in the same plane, this
has a sinusoidal dependence on the angle between the bar and the
interferometer's ``x arm'' \eqref{eq:gamma0ifobar}.  For correlations
between two bars, there is still a sinusoidal dependence on the angle
between the two bars \eqref{eq:gamma0bars}, but now that modulation is
offset by $\frac{1}{3}$, so that cross-correlation measurements with
parallel bars $\gamma(0)=\frac{4}{3}$ are more efficient than those
with perpendicular bars $\gamma(0)=-\frac{2}{3}$.  This can be
understood, as explained in Sec.~\ref{sss:modulation-DC-barbar}, in
terms of the distribution of optimal propagation directions for bar
detectors.

At higher frequencies, we have confirmed the empirical result seen in
\cite{Finn:2001} of a sinusoidal azimuthal modulation of the overlap
reduction function, generally offset from zero mean
\eqref{eq:gammafinal}.  We have also verified that the modulation has
a period of $180^\circ$ with no higher-order moments, and give
explicit expressions for the amplitude $\gamma_A(f)$, offset,
$\gamma_0(f)$ and phase $\zeta_{\text{max}}(f)$ of the azimuthal
modulation, in terms of geometrical quantities such as the detector
response tensor and the unit vectors associated with the bar
detector's location.

Finally, we have illustrated these results in
Section~\eqref{ss:modulation-examples} by applying them to several
pairs of real-world detectors.  LLO-ALLEGRO has the best observing
geometry and the greatest opportunity for modulation, with
$\gamma(900\un{Hz})$ varying between $-0.90$ and $0.96$ with azimuth.
The AURIGA bar has a reasonable observing geometry with Virgo, but
little opportunity for modulation if it were rotated, with
$\gamma(900\un{Hz})$ ranging from $0.14$ to $0.22$, while NAUTILUS and
Virgo have a slightly less favorable geometry on average, but greater
potential modulation, with $\gamma(900\un{Hz})$ between $-0.18$ and
$0.07$.

\acknowledgments

The author thanks the LIGO project, the LIGO Scientific Collaboration
and the ALLEGRO project, specifically Joe Romano, Nelson Christensen,
Sam Finn, Albert
Lazzarini, Bruce Allen, Sukanta Bose, Martin McHugh, and Warren
Johnson, as well as the Virgo/EGO project and the
Ist and IInd Virgo-SIGRAV Summer
Schools, the AURIGA project and the Universit\`{a} di Trento,
especially Sylvia Poggi, Lucio Baggio, Giovanni Prodi, and Stefano
Vitale, the organizers of the Eighth Gravitational Wave Data Analysis
Workshop, the Center for Gravitational Wave Astronomy at the
University of Texas at Brownsville and Texas Southmost College, and
the Albert Einstein Institute in Golm.

This work was supported by the National Science Foundation under grant
PHY-0300609.

\appendix

\section{Conventions and Definitions}
\label{app:notat}

Lowercase Latin letters ($a,b,c,\ldots$) will represent spatial
indices ranging from 1 to 3, which can be raised or lowered with the
Kronecker delta $\delta_{ab}$.  The Einstein summation convention will
be used for these indices.

The propagation direction of a plane wave is defined by a unit vector
$\hat{n}$; given such a unit vector, one can choose an arbitrary unit
vector $\hat{\ell}$ perpendicular to $\hat{n}$ and then define
$\hat{m}=\hat{n}\times\hat{\ell}$ so that
$\{\hat{\ell},\hat{m},\hat{n}\}$ form a right-handed orthonormal
basis.  Given such a basis, it is conventional to define transverse,
traceless polarization tensors
\begin{subequations}
  \begin{align}
    e_{+}^{ab}(\hat{n}) &= \ell^a \ell^b - m^a m^b \\
    e_{\times}^{ab}(\hat{n}) &= \ell^a m^b + m^a \ell^b
  \end{align}
\end{subequations}
Note that while these tensors span the space of traceless, symmetric
tensors transverse to $\hat{n}$, they are not orthonormal
under the obvious choice of inner product on that space, since
\begin{equation}
  \label{eq:polbasisnorm}
  e_{A\,ab}(\hat{n})\,e_{A'}^{ab}(\hat{n}) = 2\,\delta_{AA'}
\end{equation}
where $A$ and $A'$ are polarization indices, which can take on the
values $+$ and $\times$.

It is useful to define the projector onto traceless, symmetric tensors
transverse to the unit vector $\hat{n}$:
\begin{equation}
  \label{eq:TTproj}
  P^{\text{TT}\hat{n}}{}^{ab}_{cd} 
  = \frac{1}{2}\sum_{A=+,\times} e_{A}^{ab}(\hat{n})\,e_{A\,cd}(\hat{n})
\end{equation}
The factor of $1/2$ is needed to make this a projector, because of the
factor of two appearing in \eqref{eq:polbasisnorm}.  Note that while
the polarization basis tensors depend on the (arbitrary) choice of
$\hat{\ell}$, the projector only depends on the unit vector $\hat{n}$.
Note also that, because it is a projector onto a two-dimensional
subspace,
\begin{equation}
  P^{\text{TT}\hat{n}}{}^{ab}_{ab} = 2
\end{equation}

Another useful projector is that onto traceless, symmetric tensors:
\begin{equation}
  \label{eq:PTTtrace}
  P^{\text{T}}{}^{ab}_{cd} 
  = \delta_{(c}^a\delta_{d)}^b - \frac{1}{3}\delta^{ab}\delta_{cd}
\end{equation}
Note that because this is a projector onto a five-dimensional subspace,
\begin{equation}
  \label{eq:PTtrace}
  P^{\text{T}}{}^{ab}_{ab} = 5
\end{equation}

Finally, we note here the forms of the response tensors for
interferometric and resonant bar detectors.  The response of a
gravitational wave detector to a GW-induced metric perturbation
$h_{ab}(t,\vec{r})$ is
\begin{equation}
  h(t) = h_{ab}(t,\vec{r}_{\text{det}})\,d^{ab}
\end{equation}
where $\vec{r}_{\text{det}}$ is the spatial location of the detector
and $d^{ab}$ is a symmetric tensor describing the detector geometry.
The response tensor for an IFO with arms along the (not
necessarily perpendicular) unit vectors $\hat{x}$ and $\hat{y}$ is
\begin{equation}
  \label{eq:difo}
  d_{\text{(ifo)}}^{ab} = \frac{1}{2}(\hat{x}^a \hat{x}^b - \hat{y}^a \hat{y}^b)
\end{equation}
while that for a cylindrical resonant bar detector with its long axis
along the unit vector $\hat{u}$ is
\begin{equation}
  \label{eq:dbar}
  d_{\text{(bar)}}^{ab} = \hat{u}^a \hat{u}^b
\end{equation}
Note that we choose not to make the response tensor for the bar
detector explicitly traceless.  It is often useful to work with the
traceless piece of a response tensor
\begin{equation}
  d^{\text{T}}{}^{ab} = P^{\text{T}}{}^{ab}_{cd}\,d^{cd}
  = d^{ab} - \frac{1}{3}\delta^{ab} d^c_c
\end{equation}
which for a bar is
\begin{equation} 
  d_{\text{(bar)}}^{\text{T}\,ab} = \hat{u}^a \hat{u}^b - \frac{1}{3}\delta^{ab}
\end{equation}
($d_{\text{(ifo)}}^{ab}$ is, of course, already traceless)

Finally, a standard pair of quantities used in describing GW detector
response is a detector's
\textit{beam pattern functions} for
the two polarization states, defined by
\begin{equation}
  F_{A}(\hat{n}) = d^{ab}\,e_{A\,ab}(\hat{n})
\end{equation}

\section{Derivation of the co\"{e}fficients in the overlap reduction function}
\label{app:deriv}

In this appendix we find the values of the co\"{e}fficients
$\rho_1(\alpha)$, $\rho_2(\alpha)$, and $\rho_3(\alpha)$ appearing in
\begin{equation}
  \label{eq:Gammaexp}
  \Gamma^{ab}_{cd}(\alpha,\hat{s})
  = \frac{5}{4\pi} \iint d^2\Omega_{\hat{n}}\,
  P^{\text{TT}\hat{n}}{}^{ab}_{cd}\,e^{i\alpha\hat{n}\cdot\hat{s}}
  = \sum_{n=1}^3 \rho_n(\alpha) T_n{}^{ab}_{cd}
\end{equation}
[cf.~\eqref{eq:Gammaint} and \eqref{eq:Gammasum}] We do this by
contracting each of the $\{T_n{}^{ab}_{cd}\}$ in turn with
$\Gamma^{cd}_{ab}(\alpha,\hat{s})$:
\begin{subequations}
  \begin{align}
    T_1{}^{ab}_{cd}\Gamma^{cd}_{ab}(\alpha,\hat{s})
    &= \frac{5}{4\pi} \iint d^2\Omega_{\hat{n}}\,
    P^{\text{TT}\hat{n}}{}^{ab}_{ab}\,e^{i\alpha\hat{n}\cdot\hat{s}}
    \\
    T_2{}^{ab}_{cd}\Gamma^{cd}_{ab}(\alpha,\hat{s})
    &= \frac{5}{4\pi} \iint d^2\Omega_{\hat{n}}\,
    P^{\text{TT}\hat{n}}{}^{ab}_{ac}\hat{s}_b \hat{s}^c
    \,e^{i\alpha\hat{n}\cdot\hat{s}}
    \\
    T_3{}^{ab}_{cd}\Gamma^{cd}_{ab}(\alpha,\hat{s})
    &= \frac{5}{4\pi} \iint d^2\Omega_{\hat{n}}\,
    P^{\text{TT}\hat{n}}{}^{ab}_{cd}\hat{s}_a \hat{s}_b \hat{s}^c \hat{s}^d
    \,e^{i\alpha\hat{n}\cdot\hat{s}}
  \end{align}
\end{subequations}
The three contractions appearing in the integrands can be written
\begin{subequations}
  \begin{align}
    P^{\text{TT}\hat{n}}{}^{ab}_{ab} &= 2 \\
    P^{\text{TT}\hat{n}}{}^{ab}_{ac} \hat{s}_b \hat{s}^c
    &= 1 - (\hat{n}\cdot\hat{s})^2 \\
    P^{\text{TT}\hat{n}}{}^{ab}_{cd} \hat{s}_a \hat{s}_b \hat{s}^c \hat{s}^d
    &= \frac{1}{2}
    \left[(e_{+}^{ab}(\hat{n})\,\hat{s}_a \hat{s}_b)^2
      + (e_{\times}^{ab}(\hat{n})\,\hat{s}_a \hat{s}_b)^2\right]
    = \frac{1}{2} \left[1-(\hat{n}\cdot\hat{s})^2\right]^2
  \end{align}
\end{subequations}
Where we have used
\begin{equation}
  P^{\text{TT}\hat{n}}{}^{ab}_{ac} = \frac{1}{2}
  \left[e_{+}^{ab}(\hat{n})\,e_{+\,ac}(\hat{n})
    + e_{\times}^{ab}(\hat{n})\,e_{\times\,ac}(\hat{n})\right]
  = \ell^b\ell_c + m^b m_c = \delta^b_c - n^a n_c
\end{equation}
and, defining a spherical co\"{o}rdinate system such that
\begin{equation}
  \hat{s} = \sin\theta\cos\phi\ \hat{\ell} + \sin\theta\sin\phi\ \hat{m}
  + \cos\theta\ \hat{n}
  \ ,
\end{equation}
\begin{equation}
  e_{+}^{ab}(\hat{n})\,\hat{s}_a \hat{s}_b =
  (\hat{\ell}\cdot\hat{s})^2 - (\hat{m}\cdot\hat{s})^2 =
  \sin^2\!\theta(\cos^2\!\phi-\sin^2\!\phi) = \sin^2\!\theta\cos 2\phi
\end{equation}
and
\begin{equation}
  e_{\times}^{ab}(\hat{n})\,\hat{s}_a \hat{s}_b =
  \hat{s}\cdot\tens{e}_{\times\hat{n}}\cdot\hat{s} =
  2(\hat{\ell}\cdot\hat{s})(\hat{m}\cdot\hat{s}) =
  2\sin^2\!\theta\cos\phi\sin\phi = \sin^2\!\theta\sin 2\phi
\end{equation}
This makes the contractions
\begin{subequations}
  \begin{align}
    T_1{}^{ab}_{cd}\Gamma^{cd}_{ab}(\alpha,\hat{s})
    &= 5\int_{-1}^1 e^{i\alpha\mu}\,d\mu
    = 10\frac{\sin\alpha}{\alpha} = 10j_0(\alpha) \\
    T_2{}^{ab}_{cd}\Gamma^{cd}_{ab}(\alpha,\hat{s})
    &= \frac{5}{2}\int_{-1}^1 (1-\mu^2) e^{i\alpha\mu}\,d\mu
    = 5\left(-\frac{2\cos\alpha}{\alpha^2}+\frac{2\sin\alpha}{\alpha^3}\right)
    = 10\frac{j_1(\alpha)}{\alpha} \\
    T_3{}^{ab}_{cd}\Gamma^{cd}_{ab}(\alpha,\hat{s})
    &= \frac{5}{4}\int_{-1}^1 (1-2\mu^2+\mu^4) e^{i\alpha\mu}\,d\mu
    = 5
    \left(
      - \frac{4\sin\alpha}{\alpha^3}
      - \frac{12\cos\alpha}{\alpha^4}
      + \frac{12\sin\alpha}{\alpha^5}    
    \right)
    = 20\frac{j_2(\alpha)}{\alpha^2}
  \end{align}
\end{subequations}

To work out the contractions of the $\{T_n{}^{ab}_{cd}\}$ with the
right-hand side of \eqref{eq:Gammaexp}, we just need to contract each
of them with the others.  We demonstrate the calculation explicitly
here in our notation:
\begin{equation}
  T_1{}^{ab}_{cd}T_1{}^{cd}_{ab} = P^{\text{T}}{}^{ab}_{ab} = 5  
\end{equation}
To get
\begin{equation}
  T_1{}^{ab}_{cd}T_2{}^{cd}_{ab} = P^{\text{T}}{}^{ab}_{ac} \hat{s}_b \hat{s}^c
\end{equation}
we note that
\begin{equation}
   P^{\text{T}}{}^{ab}_{ac} = \frac{1}{2}\delta^a_a \delta^b_c
   + \frac{1}{2}\delta^a_c\delta^b_a - \frac{1}{3}\delta^{ab}\delta_{ac}
   =
   \left(
     \frac{3}{2} + \frac{1}{2} - \frac{1}{3}
   \right)\delta^b_c
\end{equation}
so
\begin{equation}
  T_1{}^{ab}_{cd}T_2{}^{cd}_{ab} = \frac{5}{3}
\end{equation}
Next
\begin{equation}
  T_1{}^{ab}_{cd}T_3{}^{cd}_{ab}
  = P{}^{\text{T}}{}^{ab}_{cd} \hat{s}_a \hat{s}_b \hat{s}^c \hat{s}^d
  = 1 - \frac{1}{3} = \frac{2}{3}
\end{equation}
To work out
\begin{equation}
  T_2{}^{ab}_{cd}T_2{}^{cd}_{ab} =
  P{}^{\text{T}}{}^{ab}_{ef} \hat{s}^f \hat{s}_b \hat{s}^c \hat{s}_g
  P{}^{\text{T}}{}^{eg}_{ac}
\end{equation}
We note that
\begin{equation}
  P{}^{\text{T}}{}^{ab}_{ef} \hat{s}^f \hat{s}_b
  = \frac{1}{2}\delta^a_e + \frac{1}{2} \hat{s}^a \hat{s}_e
  - \frac{1}{3} \hat{s}^a \hat{s}_e
  = \frac{1}{2}
  \left(
    \delta^a_e + \frac{1}{3} \hat{s}^a \hat{s}_e
  \right) 
\end{equation}
so
\begin{equation}
  T_2{}^{ab}_{cd}T_2{}^{cd}_{ab} =
  = \frac{1}{4}
  \left(
    \delta^a_e + \frac{1}{3} \hat{s}^a \hat{s}_e
  \right)
  \left(
    \delta^e_a + \frac{1}{3} \hat{s}^e \hat{s}_a
  \right)
  = \frac{1}{4} \left(3+\frac{1}{3}+\frac{1}{3}+\frac{1}{9}\right)
  = \frac{17}{18}
\end{equation}
To calculate
\begin{equation}
  T_2{}^{ab}_{cd}T_3{}^{cd}_{ab} = \hat{s}^a \hat{s}^b P{}^{\text{T}}{}^{ef}_{ab}
  \hat{s}_f \hat{s}^g  P{}^{\text{T}}{}^{cd}_{eg} \hat{s}^c \hat{s}^d  
\end{equation}
we note that
\begin{equation}
  \hat{s}^a \hat{s}^b P{}^{\text{T}}{}^{ef}_{ab} \hat{s}_f
  = \left(\hat{s}^e \hat{s}^f - \frac{1}{3}\delta^{ef}\right)\hat{s}_f
  = \frac{2}{3}\hat{s}_e
\end{equation}
so
\begin{equation}
  T_2{}^{ab}_{cd}T_3{}^{cd}_{ab} = \frac{4}{9}
\end{equation}
Finally,
\begin{equation}
  T_3{}^{ab}_{cd}T_3{}^{cd}_{ab} = 
  \left[
    \left(\hat{s}_a \hat{s}_b - \frac{1}{3}\delta_{ab}\right)
    \left(\hat{s}^a \hat{s}^b - \frac{1}{3}\delta^{ab}\right)
  \right]^2
  =
  \left(
    1 - \frac{1}{3} - \frac{1}{3} + \frac{3}{9}
  \right)^2 = \frac{4}{9}
\end{equation}
We can summarize these results as 
\begin{equation}
  \begin{pmatrix}
    T_1{}^{ab}_{cd} \\ T_2{}^{ab}_{cd} \\ T_3{}^{ab}_{cd}
  \end{pmatrix}
  \begin{pmatrix}
    T_1{}^{ab}_{cd} & T_2{}^{ab}_{cd} & T_3{}^{ab}_{cd}
  \end{pmatrix}
  =
  \begin{pmatrix}
    5 & 5/3 & 2/3 \\
    5/3 & 17/18 & 4/9 \\
    2/3 & 4/9 & 4/9
  \end{pmatrix}
\end{equation}
which we can use, along with
\begin{equation}
  \Gamma^{cd}_{ab}(\alpha,\hat{s})
  =
  \begin{pmatrix}
    T_1{}^{ab}_{cd} & T_2{}^{ab}_{cd} & T_3{}^{ab}_{cd}
  \end{pmatrix}
  \begin{pmatrix}
    \rho_1(\alpha) \\ \rho_2(\alpha) \\ \rho_3(\alpha)
  \end{pmatrix}
\end{equation}
to combine the three sets of contractions in the
matrix equation
\begin{equation}
  \begin{pmatrix}
    10 j_0(\alpha) \\
    10 \frac{j_1(\alpha)}{\alpha} \\
    20 \frac{j_2(\alpha)}{\alpha^2}
  \end{pmatrix}
  =
  \begin{pmatrix}
    T_1{}^{ab}_{cd} \\ T_2{}^{ab}_{cd} \\ T_3{}^{ab}_{cd}
  \end{pmatrix}
  \Gamma^{cd}_{ab}(\alpha,\hat{s})
  =
  \begin{pmatrix}
    5 & 5/3 & 2/3 \\
    5/3 & 17/18 & 4/9 \\
    2/3 & 4/9 & 4/9
  \end{pmatrix}
  \begin{pmatrix}
    \rho_1(\alpha) \\ \rho_2(\alpha) \\ \rho_3(\alpha)
  \end{pmatrix}
\end{equation}
Inverting the matrix gives
\begin{equation}
  \begin{pmatrix}
    \rho_1(\alpha) \\ \rho_2(\alpha) \\ \rho_3(\alpha)
  \end{pmatrix}
  =
  \begin{pmatrix}
    \frac{1}{2} & -1 & \frac{1}{4} \\
    -1 & 4 & -\frac{5}{2} \\
    \frac{1}{4} & -\frac{5}{2} & \frac{35}{8}
  \end{pmatrix}
  \begin{pmatrix}
    10 j_0(\alpha) \\
    10 \frac{j_1(\alpha)}{\alpha} \\
    20 \frac{j_2(\alpha)}{\alpha^2}
  \end{pmatrix}
  =
  \begin{pmatrix}
    5 & -10 & 5 \\
    -10 & 40 & 50 \\
    \frac{5}{2} & -25 & \frac{175}{2}
  \end{pmatrix}
  \begin{pmatrix}
    j_0(\alpha) \\
    \frac{j_1(\alpha)}{\alpha} \\
    \frac{j_2(\alpha)}{\alpha^2}
  \end{pmatrix}
\end{equation}
which are the standard co\"{e}fficients in the expansion
\eqref{eq:gammaresult}\cite{Allen:1999}.


\begin{thebibliography}{9}
\bibitem{Christensen:1992} 
  Christensen N L 1992 \PR {\bf D46} 5250.
\bibitem{Allen:1997}
  Allen B 1997
  in \textit{Proceedings of the Les Houches School on Astrophysical Sources of 
  Gravitational Waves}, 
  eds Marck J A and Lasota J P, Cambridge, 373;
  e-Print: gr-qc/9604033.
\bibitem{Maggiore:2000} Maggiore M 2000 \PRpt {\bf 331} 28;
  e-Print: gr-qc/9909001
\bibitem{Mauceli:1996} Mauceli E et al 1996 \PR {\bf D54} 1264;
  e-Print: gr-qc/9609058
\bibitem{Finn:2001} Finn L S and Lazzarini A \PR {\bf D64} 082002;
  e-Print: gr-qc/0104040
\bibitem{Whelan:2003}
  Whelan J T et al 2003 \CQG {\bf 20} S689;
  e-Print: gr-qc/0308045
\bibitem{Whelan:2005}
  Whelan J T et al 2005 \CQG {\bf 22} S1087;
  e-Print: gr-qc/0506025
\bibitem{Abbott:2006}
  Abbott B et al (LIGO Scientific Collaboration) 2006 in preparation
\bibitem{Allen:1999}
  Allen B and Romano J D 1999 \PR {\bf D59} 102001;
  e-Print: gr-qc/9710117.
\bibitem{Flanagan:1993}
  Flanagan \'{E} \'{E} 1993 \PR {\bf D48} 2389;
  e-Print: astro-ph/9305029
\bibitem{Michelson:1987}
  Michelson P F 1987 \MNRAS {\bf 227} 933;
\bibitem{Christensen:1997}
  Christensen N L 1997 \PR {\bf D55} 448.
\bibitem{detgeom}
  http://lscsoft.ligo.caltech.edu/ligotools/detgeom/
\bibitem{Astone:2003}
  Astone P et al (IGEC Collaboration) \PR {\bf D68} 022001;
  e-Print: astro-ph/0302482
\end{thebibliography}
\end{document}